\documentstyle[12pt,aasms4,epsf]{article}

\def\ee #1 {\times 10^{#1}}
\def\ut #1 #2 { \, \hbox{#1}^{#2}}
\def\u #1 { \, \hbox{#1}}

\def\msol{\, \hbox{$\hbox{M}_\odot$}}
\def\kms {\, \hbox{km}\,\hbox{s}^{-1}}

\def\percc {\, \hbox{cm}^{-3}}

\let\grad=\nabla
\def\cross{{\bf \times}}
\def\curl #1 {\grad \cross #1}
\def\div #1 {\grad \cdot #1}

\tighten
\singlespace

\def\msol   {\hbox{$M_\odot$}}                  
\def\kms    {\hbox{km{\hskip0.1em}s$^{-1}$}}    
\def\etal   {{\it et al. }}                     

\begin{document}

\title{The Origin of the Galactic Center Nonthermal Radio Filaments: 
Young Stellar Clusters}
\author{F. Yusef-Zadeh}
\affil{Department of Physics and Astronomy, Northwestern University,
Evanston, IL. 60208 (zadeh@northwestern.edu)}

\begin{abstract}

The unusual class of magnetized nonthermal radio filaments (NTF), threads
and
streaks with their unique physical characteristics are found only within
the inner couple of degrees of the Galactic center.  Also, a number of
young, mass-losing and rare stellar clusters are recognized to lie in the
Galactic center region.  The latter characteristic of the Galactic center
region is used to explain the origin of the nonthermal radio filaments.  We
consider a mechanism in which the collective winds of massive WR and OB 
stars  within  a dense stellar environment produce shock waves that 
can  accelerate particles to
relativistic energies.  This mechanism is an extension of a
model
originally proposed by Rosner and Bodo (1996),  who suggested that
energetic
nonthermal particles are produced in a terminal shock of mass-losing
stars. 
The large-scale distribution of
the magnetic field in the context of this model  is argued to  have
neither poloidal geometry
nor pervasive  throughout the Galactic center region.

\end{abstract}

\section{Introduction}

\noindent Over the last two decades radio continuum observations
 of the Galactic
center region have revealed a large number of systems of nonthermal
radio filaments (NRFs) or nonthermal filaments (NTF)~\footnote{
NRF are used to distinguish them from nonthermal X-ray filaments}
within the inner two degrees of the Galactic 
center
(e.g., Yusef-Zadeh, Morris and Chance 1984;  Liszt 1985; Morris and
Yusef-Zadeh 1985; Bally and Yusef-Zadeh 1989;  Gray et al. 1991;  
Anantharamaiah et al 1991;   Lang et al. 1999a;  LaRosa et al. 2000). 
Some of the general characteristics of the filaments are as follows: 

\begin{enumerate}

\item The
transverse  dimensions of the long NRFs are roughly a fraction of a pc at 
the
Galactic center distance of 8.5 kpc and their length is of the order 
tens of parsecs. 

\item Most of the long and bright filaments are aligned to within 
about
30$^0$ of the rotation axis of the Galaxy but recently some have been
found to be running parallel to the Galactic plane.  The short 
and faint filamentary structures, 
known  as the ``streaks'', with lengths less than a   few parsecs 
do not appear to be preferentially oriented  perpendicular
to the Galactic 
plane.

\item  Many of 
the individual filaments or the
so-called ``threads'' break up into  multiple (at least two) 
subfilaments
that flare at their endpoints. Some filaments show a gentle curvature 
and kinks  along their lengths. The brightness of some filaments 
peak in midpoints as they gently curve.  

\item The  combination of strongly linearly polarized emission from NRFs
and  radio spectral index distribution suggest a nonthermal
synchrotron origin. 

\item 
The rotation measure (RM)  values of the NRFs range 
between a few hundred to several thousand  rad. m$^{-2}$ and the
polarization
measurements indicate that the NRFs trace the magnetic field with a
equipartition strength ranging between several to hundreds of
micro-Gauss.  

\item The NRFs show a wide range of spectral index values based on radio 
continuum observations.
Some filaments may occur in isolation with a steep
spectral index or may be part of a network of parallel filaments with a
relatively flat or inverted spectral index. A number of them show a
steepening of the spectral index at higher frequencies between $\lambda$6 and 2cm.
Since the NRFs are fairly extended and 
interferometric measurements using different array configurations 
have  different surface brightness 
sensitivity to extended emission,  
the  spectral index measurements could suffer from  this  
systematic   uncertainty. 

\item
A number of NRFs appear to be located in the vicinity of star forming
regions.

\end{enumerate}

Theoretically, it has been challenging to understand the nature of these
filaments that resemble extragalactic radio jets but are not accompanied
with any 
obvious source of acceleration of charged particles to 
 high energy relativistic energies. Although a 
number of
detailed models have been considered,  there is no consensus as to 
the
origin of the nonthermal filaments (NRFs).  These models
suggest that molecular and ionized gas clouds, mass-losing stars, Galactic
winds, and magnetic activity of the massive black hole at the Galactic
center play a role in the processes that lead to the production of the NRFs
(e.g., Heyvaerts et al.  1988; Benford 1988, 1998; Morris and
Yusef-Zadeh
1989;  Serabyn and Morris 1994;  Nicholls and LeStrange 1995;  Bally
and Yusef-Zadeh 1989;  Serabyn and G\"usten 1991; Rosso \& Pelletier 1993;
Ryutov et al. 2000; Lesch and Reich 1992; Rosner and Bodo 1996;  Dahlburg
et al.  2002; Shore and LaRosa 1999; Bicknell and Li 2001a). In most
models, the magnetic field is strong and its  global geometry in the central
region of
the Galaxy is considered to be poloidal and static. However, some recent
models have argued that the magnetic field is local and dynamic (LaRosa,
Lazio and Kassim 2001).

A review of a plethora of theoretical models of NRFs can be found in a
number of publications (e.g., Yusef-Zadeh 1989; Morris 1996; Morris and
Serabyn 1996; Bicknell and Li 2001b). It is beyond the scope of this paper
to discuss the many assumptions that have been made in these models.
Observationally, systems of NRFs are known to have some generic physical
characteristics, as described above, but noteable differences between them
have also been observed.  Here, the observational properties of individual
NRFs are summerized.  We also describe HII complexes found in the vicinity
of NRFs;  their direct interactions with each other are inconclusive.  
Following the summary, we examine the origin of the NRFs and concentrate on
the idea set forth by Rosner and Bodo (1996), who suggested that
mass-losing stellar sources are responsible for accelerating nonthermal
particles. We expand this idea and argue that the nonthermal emission from
the Galactic center filaments originates from the shocked region of the
colliding stellar winds of young clusters 	or young stellar binary 
systems in star forming regions. The model predicts
young and compact stellar clusters with multiple WR--OB binary systems
or young massive binary systems with their corresponding 
HII regions distributed in the vicinity of NRFs.

\section{Case by Case Characteristics of NRFs}

{\bf{G0.2-0.0 (Radio Arc and Its Extensions)}}: The prototype filamentary
structure of the radio continuum Arc resolves into a set of more than
a dozen
vertical filaments with lengths of about 30 pc distributed symmetrically
with respect to the Galactic equator (Yusef-Zadeh, Morris and Chance 1984;  
Yusef-Zadeh and Morris 1987a,b,c).  The radio Arc is known to be the best
example of a network of NRFs running perpendicular to the Galactic plane.
The NRFs generally show inverted spectrum $\alpha$=+0.3 (Inoue \etal 1984;
Tsuboi \etal 1985; Seiradakis et al. 1985; Sofue et al. 1989; Reich \etal
2000) where F$_{\nu}\propto \nu^{\alpha}$ with the exception of one steep
spectrum filament to the south of the Arc with $\alpha=$--0.4 between 90
and 20cm (Anantharamaiah et al. 1991). The vertical filaments of the Arc
extend toward positive and negative latitudes --0.75$^0<b<+0.75^0$ as they
become more diffuse, weaker in their surface brightness
 (Sofue and Handa 1984). High-resolution
radio continuum images show that the emission from the extensions of the
Arc is dominated by diffuse structures as well as a number of weak and
coherent filamentary features running in the direction away from the
Galactic plane (see Fig.  4b of Yusef-Zadeh 1989; Yusef-Zadeh et al. 1990;
Yusef-Zadeh and Cotton 2003, in preparation).

As the filaments of the Arc cross the Galactic equator, a dense cluster of
young and mass-losing WN or Of stars is found near G0.18-0.04.  The
Quintuplet cluster contains 10$^3$ stars and has an angular size of
$\sim25''$ (1 pc) with an age estimated to be 3-5 Myrs (e.g., Figer \etal
1999a). Prominent molecular and ionized gas clouds (G0.18-0.04 and
G0.11-0.11) are also distributed in the vicinity of the Quintuplet cluster
(Serabyn and Morris 1994; Tsuboi et al. 1997; Yusef-Zadeh, Roberts \&
Wardle 1997).

{\bf{ G0.08+0.15 (Northern Thread)}}: The isolated linearly polarized
filamentary structure which extends for about 12$'$ (30 pc) shows a curvature
in the direction away from the rotation axis of the Galaxy (Morris and
Yusef-Zadeh 1985; Yusef-Zadeh 1986; Lang, Morris and Echevarria 1999b).  
The
filament breaks up into at least two parallel components in its northwestern
extension and its brightness peaks close to its midpoint. The spectral index
value is estimated to be $\alpha$=--0.6 between 6 and 90cm but steepens to
the value of $\alpha$=--2 between 2 and 6cm (Anantharamaiah et al. 1991; Lang
et al. 1999b).  The equipartition magnetic field is estimated to be 
140$\mu$G
with a synchrotron lifetime of 4$\times10^4$ years assuming that the 
break in the spectrum of the filament occurs at 6cm (Lang et al. 1999b).  
The spectral index distribution appears to be constant along the filament.

A massive cluster of hot stars known as the Arches cluster is found in the
vicinity of the terminus of G0.08+0.15 closest to the Galactic plane.  This
consists of mainly 150 O star candidates with stellar masses greater than
20~M$_\odot$.  The Arches cluster is $\sim15''$ across, with an estimated
density of $3\times10^5$ \msol pc$^{-3}$ within the inner 9$''$ (0.36 pc) of
the cluster (e.g., Cotera et al. 1996; Serabyn, Shupe \& Figer 1998;
Blum 
et al. 2001). There 
is considerable  ionized and molecular material that appear to be
associated with the Arches cluster. The stellar, ionized and molecular
materials are all distributed in the vicinity of the southern end of the
Northern Thread (Yusef-Zadeh 1986;  Serabyn and G\"usten 1991).

{\bf{ G359.96+0.09 (Southern Thread)}}: Similar to the Northern Thread
in its
extent and its morphology except that it is running 
within 12$^0$ of the rotation axis of the Galaxy without any 
evidence of  curvature.  The
endpoint of the filament closest to the Galactic plane lies about 30$''$ north
of two HII regions known as H1 and H2 which are thought to be excited by
O6 and
O7 ZAMS, respectively (Yusef-Zadeh \& Morris 1987c; Zhao et al. 1993). 
Recent near-IR observation  detected an emission line star
at the peak of H2 (Cotera et al. 1999).  
Also, 
IRS 16 which is known to be a
massive cluster of hot stars at the Galactic center,  is located about
four
arcminutes southeast of the endpoint of the filament.  Similar to the 
 Arches and the Quintuplet clusters, the IRS 16 cluster  is also 
associated with molecular and ionized
gas clouds at the Galactic center (e.g., Genzel, Hollenbach and Townes 
1994).

{\bf{G359.43-0.09 (Sgr C)}}: One of the brightest radio continuum sources
near the Galactic center,  Sgr C,  resolves into multiple filaments and
a circular
HII region (Liszt 1992; Liszt and Spiker 1995).  The filaments extend for
10$'$ toward positive latitudes with a spectral index $\alpha$=--0.5 
between
90 and 20cm (LaRosa et al. 2000).  The filaments appear to end 
abruptly inside a
molecular cloud HII complex 
 with a velocity of --65 \kms (Liszt 
1992;
Liszt and Spiker 1995).  This HII complex is also known to coincide with a
source of infrared emission at 350 $\mu$m (Hunter et al. 2000).  An O5.5V
star is suggested to be responsible for ionizing the thermal component of
Sgr C.

{\bf{G359.54+0.18 (Ripple)}}: The ripple filament is another isolated
filamentary structure that resolves into  multiple parallel components
with a
terminus that flares in the direction toward the Galactic plane.  The
brightness of the filaments peak at the midpoint where subfilaments lie
closest to each other (Bally and Yusef-Zadeh 1989).  The spectral index
is
$\alpha$=--0.8 between 90 and 20cm (Anantharamaiah et al. 1991).  Linear
polarization measurements show that the magnetic field traces the filament
(Yusef-Zadeh, Wardle and Parastaran 1997). Recent Chandra observations show
X-ray emission from the northern filament of G359.54+0.18 (Lu, Wang
and Lang 2003). A
dense molecular cloud and an HII region  are observed at the interface of 
the
eastern edge of the filaments before they deviate from a straight line 
and flare  
(Staghun et al. 1998).

{\bf{Streaks}}: A number of small scale linear filaments or the so-called
``streaks"  are observed throughout the region in the Northern and
Southern
threads, the southern extension of Sgr C and Sgr A. These features are very
similar to the long NRFs of the Arc but are shorter 
with a
length ranging between 1$'-5'$. The surface brightness of the
streaks is typically five to ten times fainter than the long 
filaments and there is  no sign of 
bending.  It is not uncommon to observe  streaks
having orientations very different than the general direction of the
prominent NRFs (Yusef-Zadeh and Cotton 2003, in preparation).  The
polarization and
spectral
index estimates of these faint features have not been determined. The
terminus of some of these filaments appears to end at a compact
circular-like
HII region.  G0.02+0.04 is an excellent  example of a streak that ends inside
the HII
region H4,  which is known to be excited by a O8.5 ZAMS (Lang et al.
1999b;
Zhao et al.  1993).

{\bf{G359.1-0.2 (Snake)}}:  Perhaps the most striking example of isolated 
NRFs
is the ``Snake'' (Gray et al.
1991, 1995). The morphology of the Snake is distinguished somewhat from 
other
Galactic center filaments by its narrow ($<10''$) width, its long
($\approx20'$) extent, and by two uncharacteristic kinks along its length.  
In addition, 
the Snake is unusual in that it shows a gradient in spectral index at
the location of the kinks (Gray et al. 1995).  The spectral index along
the
filament is generally constant and flat between 6 and 20cm with the exception
of the major kink where the spectrum steepens to $\alpha$=--0.5.  
Subfilamentation has also been detected in the vicinity of the kinks
(Gray 
et
al. 1995). The equipartition magnetic field is estimated to be 88 $\mu$G with
a synchrotron lifetime of $\approx8\times10^5$ years (Gray et al. 1995).

Toward the northern end of the Snake near the Galactic 
plane,
 there is a cluster of HII knots  near G359.16-0.04 (Caswell and Haynes 
1987).  
The filament ends in the HII knots and morphological arguments have been
used to 
associate the Snake with the HII complex 
(Uchida et al. 1996).

{\bf{G358.85+0.47 (Pelican)}}: This linearly polarized feature
extending for 7$'$ in its length, unlike any other prominent NRFs, runs
along the Galactic
plane (LaRosa et al. 2001;  Lang et al. 1999a).  The orientation of the
magnetic field follows the linear filament,  which consists of two
subfilaments
that flare at their ends.  The angular separation of G358.85+0.47 from the
Galactic center is 1.3$^0$,  which places the angular distance of this
source 
furthest from the
Galactic equator when compared with other NRFs. The spectral index 
between 90 and 20cm is $\alpha$=--0.8
 and steepens to --1.5 between 6 and 20cm (LaRosa et 
al.
2000; Lang et al. 1999).

{\bf{G359.85+0.39}}: This new system of isolated NRFs shows subfilaments
and flaring at an angular distance of 0.5$^0$ from the Galactic center 
(LaRosa et al. 2001). 
Unlike other systems of NRFs, with the exception of the Snake, 
G359.85+0.39 displays 
a gradient in its spectral index distribution
(LaRosa et al. 2001). The spectral index value varies smoothly from
$\alpha$=--0.15 to --1.1 in the direction away from the Galactic plane, 
as discussed below.

{\bf{G359.79+0.17}}: Another system of NRFs showing multiple filaments and a
curvature in the direction away from the rotation axis of the Galaxy
is G359.79+0.17 (Yusef-Zadeh and Morris 1987b; Lang et al. 1999b). The
spectral 
index value
between 20 and 90cm is estimated to be $\alpha$=--0.6 (Anantharamaiah et al.  
1991).

\section{A Model: Colliding Winds of  a Stellar Cluster}

A number of theoretical models  have proposed  interstellar mechanisms 
as a means of
achieving the acceleration of particles to explain the nonthermal nature of
NRFs. Most of these models require strong, organized, large-scale,
interstellar magnetic fields. The models additionally require specific 
relative motions of  
Galactic winds, molecular clouds, HII
regions and supernova remnants. Here we expand 
upon the model
originally proposed by Rosner and Bodo (1996; RB96),   who have used a 
stellar mechanism
for the acceleration of particles to relativistic energies. In analogy to
the shock acceleration of the solar wind, RB96 proposed
that terminal shocks of
mass-losing stars are natural places  for the  acceleration of particles 
to high
energies. They considered that under strong and weak ISM magnetic fields,
the size of the wind bubble created by a mass-losing massive star
determines the transverse size of the filaments. Once the electrons are
accelerated at the wind terminal shock, they tag along the ISM field and
flow along with the Alfv\'en velocity as they radiate synchrotron
radiation.
The length of the filaments is then determined as the byproduct of
Alfv\'en
speed and synchrotron lifetime at radio frequencies. This model then implies
that mass-losing stars with fast winds are embedded within each
individual NRF.

The RB96 model can  be applied to any mass-losing stellar systems
in the Galaxy but  does  not specifically address the rarity of 
the NRFs which  are observed {\it only}  
in the Galactic center region. Here, the 
RB96
model is extended to explain the origin of NRFs by using young, 
compact stellar
clusters to accelerate particles to high energies.  We believe that 
this is
a more viable acceleration mechanism for production of  prominent NRFs such 
as the
radio Arc or Sgr C than the individual stellar wind sources.  
This implies
that the unusual population of young stellar clusters,  which are formed
only in the Galactic center region are tied to the origin of the unique
filamentary structures observed in the same region.

\subsection{Unusual Stellar Clusters Near the Galactic Center}

As noted above, it appears that many of the prominent nonthermal
filamentary systems are  
morphologically associated with star forming regions.  Stellar clusters near
the Galactic center are a record of the history of unusual star formation
in this
unique region. The association of the NRFs to star forming sites has in
fact been argued previously but in a different context (e.g., Serabyn
and
Morris 1994; Morris and Serabyn 1996).  These authors argue that the
acceleration of relativistic particles is due to the reconnection of the
magnetic fields at the ionized surface of molecular clouds in star forming
regions. A necessary condition for the acceleration
at the cloud surface is that the cloud has  to have a relatively large
velocity with respect to an interstellar medium  which itself is 
threaded by large-scale organized magnetic field. Also, this
model assumes that
the
poloidal component of the magnetic field dominates the global geometry of the
field
in the ISM of the Galactic center.

At present, three young ($<$ 20 Myrs) clusters have been discovered within
a projected distance of 35 pc of the center of the Galaxy -- the IRS 16,
the Arches and the Quintuplet clusters. Two other sources, namely 
the Sgr A East HII regions and the H1-H8 HII regions   
appear to be 
associated with emission line stars (Cotera et al. 1999).   
Additional  young stellar clusters  are difficult 
to detect by infrared techniques due to the large differential
extinction 
toward
the Galactic center and  due to the source confusion in near-IR
wavelengths.  For example,  the 20 \kms GMC
(M--0.13--0.08) which is known to lie near the Galactic
center has a column density of $\approx10^{24}$ which corresponds to 
a visual extinction of $\approx$430 magnitudes (Coil and Ho 1999). 
Thus, the total number of embedded young clusters in the Galactic
center region such as the Arches
cluster is very uncertain (Figer et al. 2002).    
Observationally, a systematic search has recently been
conducted to find extended near-IR sources and X-ray sources resembling
the spectra of young stellar cluster candidates using the 2MASS and
Chandra surveys of the nuclear bulge of the Galaxy (Dutra and Bica 2000,
2001; Law and Yusef-Zadeh 2003). Dutra and Bica (2000) find a total of 58
star cluster candidates within the projected distance of 600 pc from the
Galactic center.

Our motivation to investigate the nature of NRFs and associate them with
star forming regions come from recent finding of massive young clusters in
the Galactic center region. All the known stellar clusters within the
inner 50 pc show emission line stars and are known to be associated with
thermal, ionized and molecular gas clouds (e.g., Nagata et al. 1995;
Cotera et al. 1996, 1999; Serabyn et al. 1998, Figer et al. 1999b; Krabbe
et al. 1991).  It has been suggested that massive stars might have
 preferentially formed in this region (Morris 
and
Serabyn 1996) and that the initial mass function of one of the young
stellar
clusters, the Arches cluster, is flat (Figer et al. 1999b).  The formation
of a number of detected young clusters in this region is not
that unusual since 
the initial conditions for star formation in the nucleus of our Galaxy 
is 
different than those found elsewhere in the Galaxy.  For example, it 
is well known that
the temperature, pressure, velocity dispersion of the population of
molecular clouds as well as the turbulent pressure of ionized medium are
much larger in the inner 200 pcs of the Galaxy than in the Galactic disk
(e.g., Morris and Serabyn 1996 and the references therein).  In addition,
the 
gas clouds
experience an unusually high tidal field in the environment of the
Galactic center (Bally et al. 1988). Theoretically,
Portegies Zwart et al. (2001, 2002) carried out numerical simulations of
the evolution of massive star clusters within $\sim$200 pc of the Galactic
center.  These simulations include the effects of stellar evolution,
physical collisions for individual and binary stars as well as Galactic
tidal field.  They conclude that the tidal dissolution time of a cluster
is about 70 Myrs but because of the crowding of stars near the Galactic
center, their projected densities drop below the background density within
about 20 Myrs. Using this selection effect, these authors predict that
the inner 200 pc of the Galaxy could harbor some 10 to 50 young star
clusters similar to the Arches and the Quintuplet clusters. The expected
high number of compact clusters with a core radius less than a pc is
expected only in the inner 200 pc of the Galaxy. This is because the
clusters in this region have to be compact in order not to be tidally
disrupted and young because of their short relaxation   time (e.g., Kim, 
Morris and Lee 1999; Portegies
Zwart et al. 2001). Similar reasoning is used to explain the high 
pressure, high density molecular gas distributed throughout the Galactic 
center region.

\subsection{Nonthermal Emission from Colliding Winds}

Additional motivation for a physical  relationship between young clusters 
and the NRFs came from a recent discovery and successful modeling of
relativistic particles generated within  young binary systems. 
It  is well known that WR and OB stars lose strong ionized winds as
they
emit thermal radio continuum emission from an optically thick surface
located at a distance hundreds of stellar radii away (Wright and Barlow
1975;  Panagia and Felli 1975; see the review by G\"udel 2002).  In the
case of binary systems of massive stars, the thermal emission from ionized
wind can be enhanced by contributions from shocked stellar winds (Stevens
1995). More recently, it has been shown that OB stars and up to 50\% of WR
stars show signatures of nonthermal synchrotron emission from regions
beyond the optically thick surface of thermal emission (Leithere et al.
1995; Chapman et al. 1999; Dougherty \& Williams 2000).  The colliding
wind model of synchrotron emission was confirmed by Dougherty and Williams
(2000),  who showed  evidence that most nonthermal WR systems  are 
binaries.
Theoretical work to explain the generation of synchrotron radio emission
involves first order Fermi acceleration in shocks within the stellar
winds (e.g., Bell 1978). At the contact discontinuity where the
 winds of a binary system collide with each other, particles are
accelerated resulting in significant radiation (Eicher and Usov 1993).  

Considering that the densest known young clusters of WR and OB stars in the
Galaxy are distributed in the Galactic center region, it is natural to
consider nonthermal emission arising from the collection of WR stars 
or  in
young stellar clusters. 
  The near-IR spectral type of stars in these
clusters is consistent with ionized stellar winds arising from
mass-losing
WN and/or Of stars with mass-loss rates $\approx (1-20)\times10^{-5}$ \msol\
yr$^{-1}$;  lower limits to the terminal velocities of the
winds range between
800 and 1200 \kms (Cotera \etal 1996). The colliding thermal winds of the
Galactic center clusters have also been proposed to explain 
the detection of X-rays from the Arches and the Quintuplet 
clusters (Yusef-Zadeh et al. 2002; Law and Yusef-Zadeh 2003).
If indeed thermal X-rays are the result of colliding winds, previous studies
of 30 Doradus support the idea  that the binary fraction in a young
compact cluster
is extremely high (Portegies Zwart, Pooley and Lewin  2002).

Although the evidence for thermal and nonthermal emission from
individual mass-losing stars 
is well
known, the nonthermal characteristics of WR and OB stars in a
young, compact cluster environment have not been studied extensively. 
The nonthermal emission from the shocked region of the colliding
winds is believed to result  from first order Fermi acceleration
which could arise from
young stellar clusters. 
Ozernoy, Genzel and Usov (1996) have 
pointed out that the conditions that 
are 
necessary for 
diffusive shock acceleration are met  by shocks in the colliding
winds at the 
stellar core. Diffuse and compact  nonthermal emission could arise 
from the  contributions of three components: the individual 
tight binaries in the
cluster,  the   colliding winds from any two nearby massive stars within
the cluster, and 
the collision between
the hot thermal  cluster flow generated from an ensemble of 
colliding winds with  the ISM.  Each of these components is
described below. 

Based on X-ray observations, binary
systems are expected to  populate heavily  the   compact young  star 
clusters (e.g., Portegies Zwart et al. 2002). 
Radio observations by Dougherty and Williams (2000) show 
evidence of  nonthermal emission from up to 60\% in their sample of
WR stars. Radio luminosity (L$_R$) of typical WR stars which are
likely to be  
in   binary systems   is estimated to
be $\approx10^{29}$ erg s$^{-1}$ (Chapman et al.  1999).  
The total nonthermal radio luminosity is  estimated to be 
 $\approx10^{31}$ erg s$^{-1}$ assuming that about 100 such massive
binary systems  are embedded within a dense and young stellar cluster.
Recent detection of nonthermal radio emission  from the 
Arches cluster at 327 MHz is consistent with this picture.  
Radio luminosity of the Arches cluster is  estimated to be
 $4\pi D^2 \nu F_\nu \approx 2.6\times10^{30}$
ergs s$^{-1}$, assuming that D is 8.5 kpc (Yusef-Zadeh et al. 2003). 
This estimate  is a lower limit to the total nonthermal radio luminosity 
of the cluster because of the uncertainty in the spectrum of the
Arches cluster 
at low frequencies. In addition, the flat spectrum of the cluster 
at high frequencies may arise from nonthermal emission generated from an
ensemble of stellar shock winds 
with a flat spectral index, as described below.  

Nonthermal emission from a young stellar cluster 
could also arise 
from the shocked zone where the winds from individual WR or OB stars
collide with each other. Because  the separation between individual
stars in the cluster is estimated to be between 10$^{16}$ and 10$^{17}$
cm, the contact discontinuity will be at a distance beyond the surface
where thermal emission is opaque to its own radiation. 
Figure 1 shows a schematic diagram of the collision of a 
stellar wind  bubble with  the shocked gas produced 
from the colliding winds of 
the remaining stars in the cluster. 
  Using the
expected theoretical value of nonthermal radio luminosity of the region
of the collision from the winds of two WR stars (Eichler and Usov 1993)  
and assuming that the mass-loss rate and the wind velocity are $\approx
4\times10^{-5}$ \msol\ yr$^{-1}$ and 1000 \kms, respectively, 
the radio luminosity  is found to be much less than $10^{31}$ erg
s$^{-1}$.
The main reason is the low value of the ratio of flow time to synchrotron
time scale 
($\eta$
in equation 16 of Eichler and Usov (1993)). This low ratio results 
from the
low value of the 
magnetic field when extrapolated from the surface of the stars. However,
this luminosity is
enhanced 
if the cluster is extremely  compact  with an average size ranging 
between 10$^{14-15}$ cm. In this case, the average separation, r,  
between stars   is small enough that the strength of the dipole magnetic
field from the surface of the star ($\propto\ \rm r^{-3}$)  is
sufficiently
high to generate radio synchrotron emission from the  diffuse shock
acceleration of electrons.

The third and perhaps the most significant contribution to the total
nonthermal emission from a young cluster could arise from the terminal
shock of a cluster flow as it escapes the core of the cluster and
encounters the ISM gas surrounding the cluster. The X-ray emitting
shock-heated gas created by the collision of individual $\sim 1000$ km/s
stellar winds in the dense cluster environment is shown to be
accelerated, attaining a flow velocity similar to the wind velocity of
individual mass-losing stellar sources at the edge of the cluster
(Cant\'o \etal 2000; Raga et al. 2002). This cluster flow is expected to
collide with the ISM gas surrounding the cluster and produces nonthermal
radio emission. The seed relativistic particles that are generated
within the binary systems of the cluster are shocked again at the
boundary of the cluster. In the process of diffuse shock acceleration, 
energetic particles moving upstream of the shock may scatter more
effectively from the strong turbulence convected with the incoming flow. 
The turbulent medium is  known to 
produce strong  scatter broadening  of radio sources toward the Galactic
center
region (e.g., Lazio and Cordes 1998). 
Assuming that the fraction of nonthermal radio
to X-ray luminosity L$_R$ /L$_{X}\approx10^{-3}$, as observed in WR
stars (Chapman et al. 1999), is the same for binary stars and young
compact stellar cluster, L$_R$ is estimated to be $\sim 10^{33}$ ergs
s$^{-1}$.  The estimated nonthermal radio luminosity of a young cluster
is within a factor of a few of the measured radio luminosity of
typical  NRFs.  In addition, the size of the core radius of a young
cluster $<$pc, where many of the electrons are accelerated to
relativistic energies, matches well with the observed lateral dimension
of typical NRFs.

\subsection{The Galactic Center Magnetic Field Strength and Geometry}

 The gas pressure in the Galactic center region is known to be high based
on a number of molecular line observations of this region (e.g. Morris and
Serabyn 1996 and the references cited therein). The magnetic field
pressure in this region is also considered to be high. However, much of
the evidence for the mG magnetic fields throughout the Galactic center
is
based
on morphological study of the NRFs and the argument that the filaments are
interacting dynamically with dense molecular clouds.  These arguments
have widely been used to support a hypothesis 
that there is a strong ordered mG magnetic
field with a poloidal geometry pervasive throughout the inner few hundred
pc of the Galaxy.
Here we examine if there is   observational  support for this widely
accepted view of the magnetic field distribution near the Galactic
center; in addition, we study 
the limit
of weak and strong ISM magnetic fields in the context of the young
cluster model. 

Once the nonthermal particles are generated, they diffuse out depending
on what the relative pressure of the ISM is to that of the stellar
cluster.  The nonthermal gas pressure of the cluster could be confined
either by the external  gas pressure or  the magnetic pressure in the immediate
vicinity of the cluster.
Alternatively, the shocked stellar wind bubble could be confined by the 
initial magnetic field that is swept up by the initial stellar outflow 
(a  more detailed discussion of this model will be given elsewhere). 
RB96  estimated the astrosphere radius   of the mass-losing star in the
case
when $\beta$ the ratio of the ISM gas to the magnetic pressure is much
greater or much less than 1.  This radius (R) which sets the transverse
dimension
of the filaments is estimated to be in the limit of $\beta << 1$ (using
eqn. (2) of RB96)

$$R (strong) =  0.035\ (M_\odot / 10^{-6} \msol\ yr^{-1} )^{0.5} \times
(v /10^3 \kms)^{0.5}
\times (B / mG)^{-1}\ pc $$

Similarly, when $\beta >> 1$, the radius is determined by 

$$R (weak) =   1.7\ (n_{0} / 1 cm^{-3})^{-1/5} \times
(L_{wind} / 10^{36}
ergs~s^{-1})^{1/5} \times (t / 10^4 yr)^{3/5}\ pc $$ 

RB96 argued that the radius of the bubble created from the mass-losing
star matches better with the transverse dimension (fraction of pc) of the
filament if the ISM magnetic field is weak.  

\subsubsection{Strong Magnetic Field }

Applying the strong limit of the magnetic field $\beta < 1$ to the cluster
model, the transverse dimension of the filament would increase by a factor
of 10. These estimates assume that the mechanical luminosity of the
cluster wind L$_{wind}$ and the mass-loss rate of the cluster are 100
times
larger than those of a typical mass-losing star.  Because of the large
mass loss rate of the cluster $\dot M_{w}\approx 10^{-4}\;\,M_\odot$
yr$^{-1}$, the ram pressure of the cluster flow can be balanced at a
radius at  a fraction of 0.35 pc by the strong (B$\sim10^{-3}$ G) ISM
magnetic field pressure; this sets the transverse dimenson of the
filaments associated with massive, young stellar clusters. This
implies the existence of large-scale pre-existing organized flux tubes
throughout the Galactic center region. 
The large scale distribution of
the magnetic field is expected to have a poloidal geometry in the
Galactic
center region. The strong magnetic field with this  geometry
has  been considered in 
a number of models explaining the origin of NRFs.
  In this hypothesis, the relativistic electrons will
illuminate the strong ISM field lines that surround the cluster.
However, apart from a large number of assumptions that have been
made, there are difficulties with this
hypothesis both on the
grounds that
there is neither direct evidence of a pervasive strong magnetic field
nor the
evidence for  poloidal geometry of the magnetic field in the Galactic
center region, as described below.  

\begin{enumerate}

\item
The streaks and G358.85+0.47 which
have orientations along the Galactic plane must lie much further away
from the Galactic center where the geometry of the field diverges from
being dipole  and the magnetic field should be weaker than the NRFs
closer to the center (Lang et al.
1999). However, the characteristics of G358.85+0.47 NRF with its 
location at a high Galactic latitude do
not appear to 
be different than typical NRFs with the exception of its orientation. 

\item 
A large number of new NRFs have recently been detected at 20 and 90cm in
the vicinity 
of prominent well-known filaments; these new NTFs show curvature 
and  orientations that differ from earlier vertical NTFs 
(Nord et al. 2002; LaRosa et al. 2002; Yusef-Zdeh, Cotton and Hewitt
2003). 

\item
The 
synchrotron lifetime ($\tau$) of a mG field requires 
large 
Alfv\'en speed ($v_a$) and   low density of ionized medium. 
For example, $\tau$ is  only 6000 years at 5 GHz, requiring a
number density of ionized gas of 0.04 cm$^{-3}$ and $v_a\sim10^4$ \kms\
to travel the 60 pc length of the Snake. 

\item
The strong magnetic
field lines of the inner 100-200 pc need to be  anchored to the plane,
presumably to the dense cores of giant molecular clouds. 

\item
The
anisotropic distribution of the structure function of the Faraday
rotation measure toward the NRF G359.5+0.8 indicates a geometry of the
magnetic field which is inconsistent with the poloidal geometry of the
field toward this source (Yusef-Zadeh, Wardle and Parastaran 1997).

\item
Zeeman measurements of OH (1720 MHz)  masers associated with
supernova remnant masers probe the magnetic field of molecular gas with
number densities ranging between 10$^{4-5} \rm cm^{-3}$. The estimate of
the magnetic field strength is close to  that
observed in
supernova remnant masers distributed outside the Galactic center region
(Brogan et al. 2000).  Additional 
Zeeman measurements of thermal OH (1665/67 MHz) were also made 
toward 13 positions of  Galactic center molecular clouds. Many 
of these clouds lie in star forming regions in the vicinity of NRFs. 
The 3$\sigma$ upper limit to 
the line of sight magnetic field is 0.3 mG (Uchida and G\"usten 1985).
This constrains the magnetic field in magnetoionized molecular
clouds 
anchoring  the vertical field lines. 

\item
A number of studies have estimated mG magnetic field along the NRFs 
by assuming that  NRFs are dynamically colliding  with
 molecular
clouds. 
 To identify the site of the interaction, a large scale
search for
OH(1720 MHz) maser emission was made over
the inner 8$^0\times1^0$
(l$\times$b) of the Galactic center (Yusef-Zadeh et al. 1999). No evidence
of maser emission 
is  found where candidate  molecular clouds
are possibly interacting with NRFs. 

\item
Lastly, the large-scale distribution of the
magnetic field inferred from dust polarization measurements have shown a
dominant component of toroidal geometry in the magnetic field
distribution among a number of dust clouds that have been mapped in the
Galactic center region (Novak et al. 2003).
\end{enumerate}


Some of the difficulties with the large-scale, organized poloidal geometry
of mG field can be resolved by envisioning a picture in which the
 magnetic field is strong but not pervasive. The ISM pressure of the
Galactic center region is non-uniformly distributed but is in pressure
equilibrium with the magnetic field pressure.  A schematic diagram of
Figure 1 shows a region where the non-thermal gas from the cluster
illuminates the strong magnetic field flux tube whose pressure is confined
by the ISM gas pressure. The narrow magnetic flux tubes lie where the ISM
and
magnetic field pressures are high.  
The
localized one-dimensional magnetic flux tubes are expected to have
a small
volume filling
factor distributed throughout the Galactic center region as they are 
expected to be surrounded by a weak magnetoionized medium. 
This implies that the  high value of the  RM distribution toward
NRFs is due to high  density of ionized material n$_e$.
 The high
rotation measure (RM)  
toward NRFs
and bright Galactic center objects is known to be due to an external  
Faraday medium. 
Assuming a typical RM$\sim$3000 rad m$^{-2}$,
as have been measured toward a number of NRFs, and a size of the Faraday
screen of 200 pc, the estimated line of sight magnetic field and electrons
density are estimated to be 2 $\mu$G and 10 cm$^{-3}$. The estimate of the
electron density and the size of the Faraday screen are also consistent
with
the  value of the emission measure 1$\times10^4$ cm$^{-6}$ pc
observed toward the Galactic center region
(Mezger and Pauls 1979; Yusef-Zadeh
et al. 1994). It is thought that the  ionized medium co-exists with
the Faraday medium and  acts as a 
scattering screen 
broadening  background compact radio sources.   
(Yusef-Zadeh et al. 1994; Lazio and Cordes 1998).

 The next question that arises in the context of the above model is why 
 most NRFs lie
perpendicular to the Galactic plane.  The non-uniform
distribution of the pre-existing flux tubes filled with strong magnetic
fields must be oriented 
perpendicular to the Galactic plane. Alternatively, it is more natural
to consider the following.  The orientation of the
prominent NRFs could be the consequence of the environment in which they
are born.  These environmental factors could preferentially 
maintain   
NRFs that run perpendicular to the Galactic plane and suppress the 
NRFs running along the Galacic plane. One selection effect 
could be due to a higher density of molecular gas distributed along the
Galactic plane than away from the plane. Giant molecular clouds with high
densities and kinetic temperatures could limit the growth of NRFs along
the equatorial plane assuming that the magnetic pressure of the filaments
is less than the molecular gas pressure in the Galactic plane of the
Galactic center region.  

The other effect is the differential rotation of
the central region of the Galaxy which is expected to distort and destroy
much of the long NRFs oriented along the Galactic plane.  The long NRFs
directed perpendicular to the Galactic plane are more likely to survive
than those oriented along the Galactic plane due to
the  ineffectiveness of the
differential rotation in the direction away from the Galactic plane. The
NRFs can survive if their $\delta r/r \le 0.1$ where $\delta$r is the
length of a linear filament when projected along the Galactic plane at a
distance r from the Galactic center.  For long filaments along the plane,
the circular trajectory of one end of the filament will be slower that the
circular velocity of the other end which is closer to the Galactic center;
thus the long filaments are dynamically distorted after a few rotations.  
This implies that the filaments along the Galactic plane must have short
lengths whereas the long filaments such as the Snake or the Arc can only
survive if they are oriented perpendicular to the Galactic plane.  As
pointed out in section 2, there does not appear to be a  trend in the
dominant orientation of
the streaks with respect to the Galactic plane.  Thus, they are not
much  affected by the above environmental factors.

\subsubsection{Weak  Magnetic Field}

The relativistic particles emerging from the cluster flow can stream along
the local magnetic field with a large value of $\beta$.  The size of the
bubble surrounding the cluster will be 17 pc if the mechanical luminosity
of the cluster L$_{wind}\approx10^{38}$ ergs s$^{-1}$ lasts for 10$^5$
years. This value of R is much larger than the size of a bubble produced
by a mass-losing star as estimated by RB96. If the density of the
surrounding medium is 10$^{5}$ cm$^{-3}$, then R will be small enough to
match the width of NRFs.  However, the estimated number density is too
high and it is unlikey that the shocked cluster flow will be collimated
when the external magnetic field is weak unless the initial magnetic field
of the cluster which is swept up by the cluster flow confines the bubble.  
We believe that when $\beta > 1$, the size of the shocked outflow from a
massive binary system instead of young clusters matches better with the
width of the filaments as
RB96 had argued. However, this scenario can account for the energetics of
the streaks and not the more luminous and prominent NRFs. Since radio
luminosity L$_R$ of the streaks is about 0.1 to 0.01 times L$_R$ of the
bright
NRFs, a WR-OB binary system could be the source of the relativistic
particles. In this scenario, a local inhomogeneity in the ISM pressure
allows the shocked gas to flow in the direction away from a binary system.
When the particle pressure is higher than the magnetic field pressure, the
nonthermal gas can diffuse along a ``channel'' that has a much lower
magnetic field and thus suffers no radiation loss. Equilibrium stability
analysis of this system has been studied in detail by Rossi et al. (1993)
where they found  that the magnetic field can be amplified by
filamentation
instability driven by synchrotron cooling provided that $\beta > 1$.
However, it is not clear how the "channel" of low magnetic field with
nonthermal gas is confined under the condition that $\beta > 1$.  Also,
the onset of instability is expected to occur typically after a
synchrotron cooling time scale which corresponds to the length of the
filaments divided by their Alfv\'en speed.  If the energy spectrum of the
relativistic particles is steep, the synchrotron cooling time could be
long. This results in a gap between the onset of the filamentary structure
and the filament origin, thus, making the hypotheis difficult to 
test observationaly.

\subsection{The Association of Young Clusters with  NRFs. 
}
\subsubsection{ The Brightness  Distribution}

The association of nonthermal radio-emitting young stellar clusters with
NRFs implies that young clusters should be embedded within every system of
prominent NRFs.  However, the dynamics of star clusters and NRFs are known
to be
different from each other during the synchrotron lifetime of NRFs.  The
circular motion of stars and gas clouds range between 100 to 200 \kms\ in
this region of the Galaxy.  The gas clouds are much more subject to
non-gravitational (i.e., tidal and magnetic) effects whereas compact young
stellar clusters are subject to the effects of the dynamical friction.  
Thus, the long NRFs may get distorted as they follow the motion of the
compact cluster.  

The motion of the cluster with respect to
the shocked bubble, which is confined either by magnetic field or by gas
pressure, distorts the symmetry at the point of origin (Weaver et al.
1977). If the space velocity of the cluster is 6 \kms, the shocked bubble
with a size of 0.3 pc becomes distorted over 10$^5$ years, the lifetime of
the outflow. Consequently, the filaments should become  broadended and 
somewhat asymmetric at the point of the origin due to the motion of the 
cluster and its shocked bubble. 
 In addition, If we assume
the relative velocity between the NRFs and the acceleration site is
between 1 and 10 \kms, the NRFs will drift by about 0.01 to 10 pc during
the synchrotron lifetime of NRFs ranging between 10$^4-10^6$ years.  
Thus, the filaments can be 
 bent at the filament origin.

Another characteristic of a number of NRFs   is that   
their  brightness peak  in the middle of the filaments.  This peak
emission does not appear to be in the vicinity of the stellar clusters
responsible for their supply of relativistic particles. In the context of
this model, we believe the deviation of the orientation of the magnetic
fields is likely to be responsible for an increase 
in the brightness of the filaments. A change in the orientation of the
magnetic field, as has been observed in  a number prominent NRFs ,
 suggests   that there are internal oblique shocks re-accelerating 
partilces to relativistic energies in midpoints where synchrotron
emissivity 
is enhanced (a more detailed account of this picture will be given
elsewhere).

\subsubsection{The Spectral Index Distribution}

The distribution of spectral index is  either steep for isolated
filaments or flat for network of filaments.  In the context of the
proposed model, the colliding winds in the core of a young cluster are
shocked multiple times before the X-ray emitting cluster flow gets
shocked again as it reaches the surrounding ISM.  Diffusive shock
acceleration by a single shock is recognized to produce a power law
energy distribution (e.g., Blandford and Eichler 1987); for a single
adiabatic shock, the expected indices are $\alpha$=--0.5.  A sequence of
identical and non-identical shocks are estimated to have an asymptotic
spectrum producing a power law with a flat spectral index $\alpha=0$
(Pope and Melrose 1994; Melrose and Pope 1993).  The spectrum due to
fast shocks evolves more rapidly toward a flat spectrum than that of
weak shocks. This implies that slower shocks have a flat spectrum over a
smaller energy range (Pope and Melrose 1994). 

Considering that the spectral index of the NRFs ranges over wide values,
the diffuse shock acceleration mechanism due to multiple shocks predicts
a flat spectrum of the synchrotron at the point of origin.  Thus, the
prediction of the model is that the origin of the NRFs should have a
flat spectrum at high radio frequencies. 
The energy  losses due
to radiation and particle escape due to diffusive effects may 
steepen the spectrum away from the filament origin.  
An additional
effect that can flatten or even invert radio spectrum is the
contribution of ionized thermal gas in star forming regions. 
The strong radiation field of young massive clusters ionize  dense
molecular clouds (e.g., the Arches and Quintuplet
clusters)  and there
should be much diffuse ionized gas in the environments from which  NRFs
are born 
(e.g.
The Arc, 
Sgr A and Sgr C). 
 The bundle
of NRFs associated with the radio Arc and Sgr C is  known to be surrounded
by
ionized thermal gas as evidenced by the detection of strong Faraday
rotation as well as the detection of 
radio recombination line emission toward  this system of nonthermal
filaments (e.g. Anantharamaiah and
Yusef-Zadeh 1989).  The spectral index values  become stepper in the
direction 
away from the Galactic plane for these sources as well as the isolated
filament G359.85+0.39 (LaRosa et al. 2001). This is consistent with the 
picture that thermal gas does not affect the intrinsic value of 
the spectral index away from the Galactic plane. 

It is possible that  thermal
ionzied gas is distributed in front of  NRFs. Alternatively,   thermal
gas with 
electron density n$_e$ cm$^{-3}$ 
are uniformly mixed with nonthermal gas along the path
length L (pc)
throughout this system of filaments. 
The apparent
synchrotron emission  in the latter situation
is given by Salter and Brown (1974) 

$$ \rm I(\nu) \propto \nu^{-\alpha+2.1} [1-exp{(-\nu_A/\nu)^{2.1}}] $$

where $\nu_A = 0.5 \times \rm n_e\times \rm L^{0.5}$  in MHz. 

At very low frequencies I$(\nu) \propto \nu^{-\alpha+2.1}$, the spectrum
is inverted if   $\alpha < $2.1. 
Low frequency observations of the 
Arc between 160 MHz and   1.4 GHz
show that the radio Arc 
has an apparent spectral index between 0.37 near
G0.16-0.15
(Yusef-Zadeh
et al. 1986).  
Considering the large uncertainty of the measured spectral index 
of G.16-0.15 using differnt spatial resolutions where there is thermal
and nonthermal emission on a wide range of angular scales, this
is 
consistent with  the inverted spectrum at low frequecies but steep   
$\alpha=$ 1.7. 
At high  frequencies, I$(\nu) \propto
\nu^{-\alpha}\times\nu_A^{2.1}$;  high resolution observations 
of the NRFs near the Arc have not detected 43 GHz emission from the 
filaments (Sofue, Murata \& Reich 1992) suggesting $\alpha >$ 0.7 which
is not
inconsistent 
with the value of $\alpha$ at low-frequencies. 
The
value of $\nu_A$ is 
estimated to be about 600 MHz corresponding to n$_e\sim400$ cm$^{-3}$
toward G0.16-0.15
if we assume that the path length L$\sim 9$ pc. 
The value of $\nu_A$ will be different along the long extent of 
the linear filemants. 

As for the spectral index of the isolated filaments, the main question 
that arises is how to account for the constant value of the
spectral index
along the filaments and yet  a steepening of $\alpha$ at higher
frequencies for a given position along the filaments. 
A break in the spectrum is interpreted to be the consequence of 
spectral aging 
of synchrotron radiation whereas the constancy of the spectral index
along the filament 
requires   shock re-acceleration along the filaments. 
This is consistent with the interpretation of  the change in the 
brightness distribution of 
of the filaments in midpoints.

\section{Conclusions}

The hypothesis outlined above supports a stellar mechanism to accelerate
particles by young dense clusters or massive binary systems using an
efficient and well known Fermi acceleration of cosmic rays.  The
population of young stellar clusters responsible for the origin of the
NRFs is considered to be unique in the Galactic center region as evidenced
by the discovery of a number of young stellar clusters (e.g., Arches
cluster).  We believe that it is not by accident that many of the
prominent NRFs are distributed in the vicinity of HII regions associated
with star forming activity. This model which is an expansion of an earlier
model by RB96 predicts that compact young stellar clusters characterized
by thermal and nonthermal emission with flat spectrum should be found in
the vicinity of individual NRFs in the Galactic center whereas massive
binary systems are responsible for the origin of the streaks which are
considered to be the scaled-down version of the prominent NRFs.

Both strong and weak  magnetic field lines in the ISM of the
Galactic center region are  considered to be 
 illuminated 
by the relativistic particles of the cluster but each has its  
own difficulties. In particular, 
we argue that 
present observations place strong constraint on the idea that 
 strong, pervasive magnetic  field with a poloidal geometry 
is distributed in the the Galactic center region. 
In the context of the 
model presented here,  we proposed  that the ISM
of
the Galactic center region has a non-uniform
distribution of strong magnetized flux tubes, 
which are confined by high gas pressure, 
but with small volume filling factor.  
We also discussed the brightness and spectral index distributions
of NRFs and concluded that shock re-acceleration of paticles 
must be taking  place along the filaments. This implies a
mechanism  which is known to operate in young stellar objects 
as  well as  extragalactic radio sources, a more detailed account 
of which  will be given elsewhere. 

Acknowledgments: I am indepted to Mark Wardle and Arieh K\"onigl  for a
number of 
illuminating discussions as well as reading of the original manuscript. 
We also thank the referee for useful comments.

\begin{figure}
\plotone{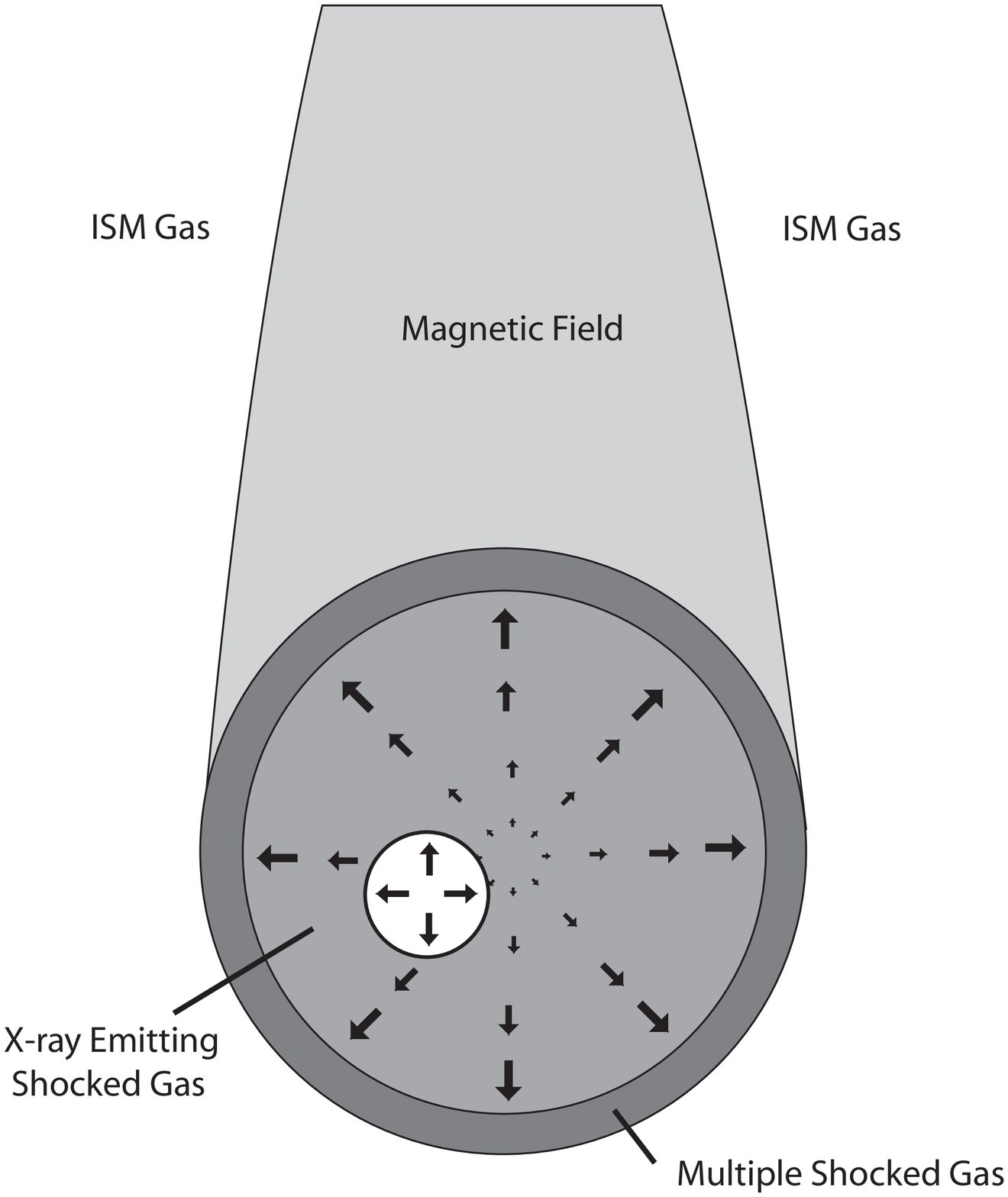}
\figcaption{A schematic diagram showing the origin of 
the relativistic particles when the winds from 
a single mass-losing stellar bubble  as well as the cluster flow 
collide with the cluster flow 
and the ISM, respectively.}
\end{figure}



\begin{references}

\reference {} Anantharamaiah, K.R., Pedlar, A., Ekers, R.D. \& Goss, W.M.
1991, MNRAS, 249, 262

\reference {} Anantharamaiah, K.R. \&  Yusef-Zadeh, F. 1989, 
in IAU Symp. 136, The Center of
the  Galaxy, ed: M. Morris (Dordrecht:Kluwer), 159

Bally, J. \& Yusef-Zadeh, F. 1989, ApJ, 336, 173

Bell, A.R. 1978, MNRAS, 182, 147


Benford, G. 1988, ApJ, 333, 735
   
Benford, G. 1998, MNRAS, 285, 573

Bicknell, G. \& Li, J. 2001a, ApJ, 548, L69

Bicknell, G. \& Li, J. 2001b, PASA, 18, 431

Blandford, R.D. \& Eichler, D. 1987, Phys. Rep., 154, 1. 
 

\reference {} Blum, R.D., Schaerer, D., Pasquali, A., Heydari-Malayeri,
M., Conti, P.S., Schultz, W. 2001, ApJ, 122, 1875

\reference {} Brogan, C.L.,  Frail, D.A.,  Goss, W.M., \& Troland, T.H.
2000, ApJ, 537, 875

\reference{} Cant\'o, J., Raga, A.C. \& Rodriguez, L.F. 2000, ApJ, 536,
896

Coil, A. \& Ho, P.T.P. 1999, ApJ, 513, 752

Chapman, J.M., Leitherer, C., Koribalski, B., Bouter, R. \& Storey, M. 
1999, 
ApJ, 518, 890


\reference {} Cotera, A., Erickson, E.F., Colgan, S., Simpson, J., Allen,
D.
\& Burton, M. 1996, ApJ, 461, 750

\reference {} Cotera, A., Simpson, J.P., Erickson, E.F. \& Colgan, S.
1999, ApJ, 510, 747


\reference {} Dahlburg, R.B., Einausdi, G., LaRosa, T.N. \& Shore, S.N. 
2002, ApJ, 568, 220

Dougherty, S.M. \& Williams, P.M. 2000, MNRAS, 319, 1005

Dougherty, S.M.,  Williams, P.M. \& Pollacco, D.L. 2000, MNRAS, 316, 143

\reference {} Dutra, C.M. \& Bica, E. 2000, A.\&A. 359, L9

\reference {} Dutra, C.M. \& Bica, E. 2001, A.\&A. 376, 434

Eichler, D. \& Usov, V. 1993, ApJ, 402, 271

\reference{} Figer, D.F.,  McLean, I.S.  \& Morris, M.,  1999, ApJ, 514,
202

\reference{} Figer, D.F., Najarro, F., Gilmore, D.,  Morris,
M., Sungwoo, S.K., Serabyn, E. et al. 2002, ApJ, 581, 275

\reference{} Figer, D.F., Sungsoo, S.K., Morris, M., Serabyn, E.,
Rich, R.M., McLean, I.S. 1999, ApJ, 525, 750

Genzel, R., Hollenbach, D. \& Townes, C.H. 1994, Rep.Prog.Phys, 57, 417


Gray, A.D., Cram, L.E., Ekers, R.D. \& Goss, W.M. 1991, Nature, 353, 237

Gray, A.D., Nicholls, J., Ekers, R.D. \& Cram, L.E. 1995, ApJ, 448, 164

G\"udel, M. 2002, ARAA, 40, 217


Heyvaerts, J., Norman, C. \& Pudritz, R.E. 1988, ApJ, 330, 718

Hunter, T.R., Churchwell, E., Watson, C., Cox, P., Benford, D.J. \& 
Roelfsma, P.R. 2000, AJ, 119, 2711

Inoue, M., Takahashi, T., Tabara, H., Kato, T. \& Tsuboi, M. 1984, PASJ,
36, 633


Kim, S.S., Morris, M. \& Lee, H.M. 1999, ApJ, 525, 228

\reference{} Krabbe, A.,  Genzel, R.,  Drapatz, S. \&
Rotacius, V. 1991, ApJ, 382,L19

LaRosa, T.N., Kassim, N.E., Lazio, T.J.W. \& Hyman, S.D. 2000, AJ, 119, 
207

LaRosa, T.N., Lazio, T.J.W. \& Kassim, N.E. 2001, ApJ, 563, 163

Lang, C., Anantharamaiah, K.R., Kassim, N.E. \& Lazio, T.J.W. 1999a, ApJ, 
521

 
Lazio, T.J.W. \& Cordes, J.M.,  1998, ApJ, 505, 715 


\reference{} Lang, C.,  Morris, M. \&  Echevarria, L. 1999b, ApJ, 526, 727

\reference{} Lang, C.C., Figer, D.F., Goss, W.M. \& Morris, M. 1999c,
AJ, 118, 2327


Lang, C., Yusef-Zadeh, F. \& Goss, W.M. 2002, in preparation.


LaRosa, T.N.,  Nord, M.E.,  Lazio, T.J.W.,  Kassim,
N.E., 2002 AAS, 201.3103L

Lazio, T.J.W. \& Cordes, J.M. 1998, ApJ, 505, 715

Law, C. \& Yusef-Zadeh, F. 2003, 
Astron. Nachr., Vol. 324, No. S1 (2003)

Leithere, C., Chapman, J. \& Koribalski, B. 1995, ApJ, 450, 289

Lesch, H. \& Reich, W. 1992, A\&A, 264, 493

Liszt, H. 1985, ApJ, 203, L65

Liszt, H.S. 1992, ApJS, 82, 495


Liszt, H.S., \& Spiker, R.W. 1995, ApJS, 98, 259

Lu, F.G., Wang, Q.D. \& Lang, C.C. 2003, AJ, submitted

Macchatto, F. et al. 1991, ApJ, 373, L55


Melrose, D.B. \& Pope, M.H. 1993, PASA, 10, 3

Mezger, P.G., \& Pauls, T. 1985, IAU Symp. 84, The Large-Scale 
Characteristics of  the Galaxy, ed: W.B. Burton (Dordrecht: Reidel), 357 


Morris, M. 1996, in IAU Symp. 169, Unsolved Problems of the Milky Way, ed. 
L. Blitz \& P.J. Teuben (Dordrecht: Kluwer), 247

Morris, M. \& Serabyn, E. 1996, ARAA, 34, 645

Morris, M. \& Yusef-Zadeh, F. 1989, ApJ, 343, 703

Morris, M. \& Yusef-Zadeh, F. 1985, AJ, 90, 2511


Nagata, T., et al. 1995, AJ, 109, 1676

Nicholls, J. \& LeStrange, E.T. 1995, ApJ, 443, 638

Nord, M. E.,  Brogan, C. L.,  Hyman, S.,  Lazio, T. J. W.,  Kassim, N. E., 
Duric, N. 2002, AAS, 201.3101N
	
Novak, G.,  Chuss, D. T.,  Renbarger, T.,  Griffin, G.S., et al. 
2003, ApJ, 583, L83

Ozernoy, L.M., Genzel, R. \& Usov, V.  1997, MNRAS, 288, 237

Owen, F.N., Hardee, P.E. \& Cornwell, T.J. 1989, ApJ, 340, 698

Panagia, N. \& Felli, M. 1975, A\&A, 39, 1

Pope, M.H. \& Melrose, D.B. 1994, PASA, 11, 175

\reference {} Portegies Zwart, S.F., Makino, J., McMillan, W. \& Hut, P.
2001, ApJ, 546, L101

\reference {} Portegies Zwart, S.F., Makino, J., McMillan, L.W., Hut, P. 
2002, ApJ,  565,  265

\reference {} Portegies Zwart, S.F., Pooley, D. \& Lewin, W.H.G., 2002, 
ApJ, 574, 762

Raga, A.C., Velazquez, P.F., Cant\'o, Masiadri, E. \& Rodriguez, L.F. 
2001, ApJ, 559 L33

Reich, W., Sofue, Y. \& Matsuo, H. 2000, PASJ, 52, 355


Rosner, R. \& Bodo, G. 1996, ApJ, 470, L49

Rossi, P., Bodo, G., Massaglia, S. \& Ferrari A. 1993, ApJ, 414, 112

Rosso, F. \& Pelletier, G. 1993, A.A. 270, 416

Ryutov, D.D., Derzon, M.S. \& Matzen, M.K. 2000, RMP, 72, 167


Salter, C.J. \& Brown, R.L. 1988, Galactic and Extragalactic Radio
Astronomy, 
ed: G.L. Verschuur and K.I. Kellerman, Springler Verlag, 1  

Shore, S.N. \& LaRosa, T.N. 99, ApJ, 521, 587

Seiradakis, J.H., Lasenby, A.N., Yusef-Zadeh, F. Wielebinski, R. \& Klein,
U. 1985, Nature, 317, 697


Serabyn, E. \& G\"usten, R.  1991, A.A.,  242, 376

Serabyn, E. \& Morris, M. 1994, ApJ, 424, L91

Serabyn, E., Shupe, D., \& Figer, D. F. 1998, Nature, 394,
448

Sofue, Y. \& Handa, 1984, Nature, 310, 568

Sofue, Y., Murata, Y. \& Reich, W. 1992, ApJ, 44, 367


Sofue, Y., Reich, W. \& Reich, P. 1989, ApJ, 341, L47


Staguhn, J., Stutzki, J., Uchida, K.I. \& Yusef-Zadeh, F. 1998, A.A., 336, 
290

Stevens, I.R. 1995, MNRAS, 277, 163

Stevens, I.R., Blondin, J.M. \& Pollack, A.M.T. 1992, ApJ, 386, 265

Tsuboi, M., Inoue, M., Handa, T., Tabara, H., Kato, T. et al. 1986, AJ,
93, 818

Tsuboi, M., M., Ukita, N., \& Handa, T. 1997, ApJ, 481,
263
Uchida, K.I. \& G\"usten, R. 1985,  A.A., 298, 473
   
Uchida, K.I., Morris, M., Serabyn, E., \& G\"usten, R. 1996, ApJ, 462, 768

Uchida, K.I., Morris, M., Bally, J., Pound, M. and Yusef-Zadeh, F.,
1992,  ApJ,  398, 128



Wright, A.E., Barlow, M.J. 1975, MNRAS, 170, 41

Yusef-Zadeh, F. 1986, PhD Thesis, 1986, Columbia University

Yusef-Zadeh, F. 1989, in IAU Symp. 136, The Center of
the  Galaxy, ed: M. Morris (Dordrecht:Kluwer), 243

\reference {} Yusef-Zadeh, F., Law, C., Wardle, M., Wang, Q. D.,
A. Fruscione \etal 2002, ApJ, 570, 665

Yusef-Zadeh, F., Morris, M. \& Chance D. 1984,
Nature, 310, 557

Yusef-Zadeh, F., Cotton, W. \& Hewitt, J. 2003, in preparation

Yusef-Zadeh, F., Cotton, W., Wardle, M., Melia, F. \& Roberts, D.A.
1994, ApJ, 434, L63

Yusef-Zadeh, F.,  Goss, W.M., Roberts, D.A.,  Robinson, B., Frail, D.A.
1999, ApJ, 527, 172


\reference {Y87} Yusef-Zadeh, F. and Morris, M. 1987a, ApJ, 320, 545

\reference {Y87} Yusef-Zadeh, F. and Morris, M. 1987b, AJ, 94, 1178

\reference {Y87} Yusef-Zadeh, F. and Morris, M. 1987c, ApJ, 329, 729

Yusef-Zadeh, F., Morris, M., Lasenby, B.A., Seiradakis, J.H. and
Wielebinski, R.  1990, in Proc.   IAU Symp. No. 140 on
Galactic and Extragalactic
Magnetic Fields, eds: R. Beck, P.P. Kronberg and R. Wielebinski, pp. 374

\reference {Y87} Yusef-Zadeh, F.,  Morris, M., Slee, O.B. \& Nelson,
G.I. 1986, ApJ, 310, 689

Yusef-Zadeh, F., Nord, M., Wardle, M., Law, C. \& Lang, C. \& Lazio, T.J.W. 
2003, ApJ, 590, L103

Yusef-Zadeh, F.,  Roberts, D. A. \&Wardle, M. 1997,  ApJ, 490, L83

Yusef-Zadeh, F., Uchida, K., \& Roberts, D.,  1995, Science, 270,
1801



Yusef-Zadeh, F., Wardle, M. \& Parastaran, P. 1997, ApJ, 475, L119

Zhao, J.H., Desai, K., Goss, W.M. \& Yusef-Zadeh, F. 1993, ApJ, 418, 235

\reference{}

\reference{} 
\reference{}
\reference{}



\end{references}
\end{document}